\documentclass{ecai} 

\usepackage{latexsym}
\usepackage{amssymb}
\usepackage{amsmath}
\usepackage{amsthm}
\usepackage{booktabs}
\usepackage{enumitem}
\usepackage{graphicx}
\usepackage{color}
\usepackage{multirow}

\newcommand{\BibTeX}{B\kern-.05em{\sc i\kern-.025em b}\kern-.08em\TeX}

\begin{document}


\begin{frontmatter}




\title{SyncGuard: Robust Audio Watermarking Capable of Countering Desynchronization Attacks}


\author[A]{\fnms{Zhenliang}~\snm{Gan}
    }

\author[A]{\fnms{Xiaoxiao}~\snm{Hu}
    }

\author[A]{\fnms{Sheng}~\snm{Li}
    }

\author[A]{\fnms{Zhenxing}~\snm{Qian}
    \thanks{Corresponding Author.}}

\author[A]{\fnms{Xinpeng}~\snm{Zhang}
    }

\address[A]{Fudan University, China}


\begin{abstract}
Audio watermarking has been widely applied in copyright protection and source tracing.
However, due to the inherent characteristics of audio signals, watermark localization and resistance to desynchronization attacks remain significant challenges.
In this paper, we propose a learning-based scheme named SyncGuard to address these challenges.
Specifically, we design a frame-wise broadcast embedding strategy to embed the watermark in arbitrary-length audio, enhancing time-independence and eliminating the need for localization during watermark extraction.
To further enhance robustness, we introduce a meticulously designed distortion layer.
Additionally, we employ dilated residual blocks in conjunction with dilated gated blocks to effectively capture multi-resolution time-frequency features.
Extensive experimental results show that SyncGuard efficiently handles variable-length audio segments, outperforms state-of-the-art methods in robustness against various attacks, and delivers superior auditory quality.
\end{abstract}

\end{frontmatter}


\section{Introduction}
\label{sec_intro}
With the booming popularity of online platforms like TikTok and Audible, sharing diverse audio creations on social media has become a prevailing trend.
These platforms have seamlessly woven into our daily routines, revolutionizing our engagement with entertainment and our quest for knowledge in profound ways.
This evolution, while transformative, introduces formidable challenges in copyright protection and the tracing of content origins.
Digital watermarking is an effective method for source tracing and copyright protection~\cite{gan2025genptw,liang2025screenmark,liu2025watermarking,liu2023dear,liu2023detecting}.
Imperceptibility and robustness represent two of the most demanding requirements in digital watermarking. Specifically, the embedded watermark should remain inaudible to human perception, while also exhibiting strong resilience, ensuring accurate recovery even after the watermarked audio undergoes unintended degradation or malicious removal attacks.

\begin{figure}[h]
  \centering
    \includegraphics[width=1.0\linewidth]{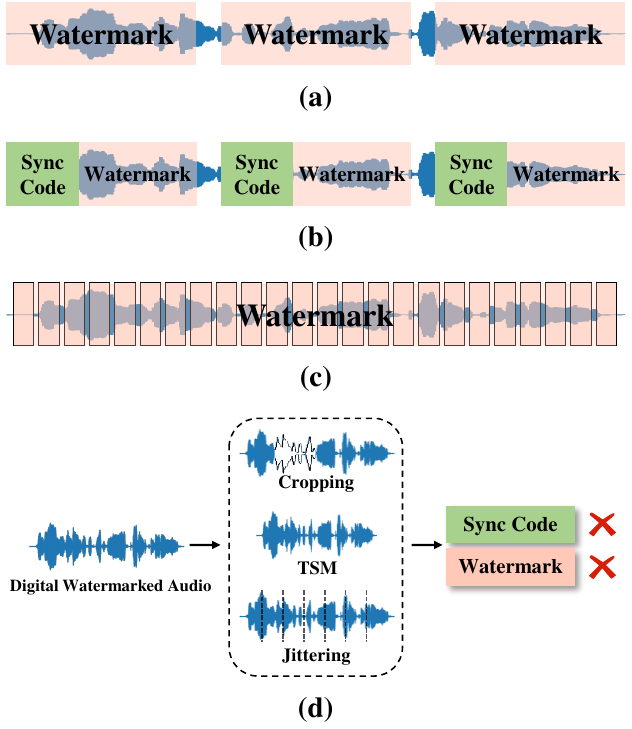}
    \vspace{-7pt}
  \caption{(a) Fixed-length embedding strategy. (b) Synchronization Code and Watermark jointly embedded. (c) Frame-wise broadcast embedding strategy of SyncGuard. (d) The illustrative diagram of desynchronization attacks.}
  \label{embedding}
\end{figure}

In practical applications, the same watermark is often repeatedly embedded at various locations within an audio segment.
This is due to the inherent nature of audio signals as functions defined along the time axis, where the length of the audio is flexible. 
During watermark extraction, the embedding location is unknown, which introduces the challenge of localization and synchronization~\cite{li2024ideaw}.
Existing methods either adopt a fixed-length embedding strategy~\cite{liu2023dear,pavlovic2022robust}, as shown in Fig.~\ref{embedding}(a), or jointly embed synchronization code and watermark information~\cite{li2024draw,li2024ideaw}, where the synchronization code is used to locate the watermark position, as illustrated in Fig.~\ref{embedding}(b).
Desynchronization attacks, such as cropping, time scale modification (TSM), and jittering, can cause changes in the watermarking location.
Although desynchronization attacks may not entirely erase all watermark information, they severely disrupt the synchronization between the embedding and decoding processes, resulting in inaccurate watermark extraction, as illustrated in Fig.~\ref{embedding}(d).
\begin{figure}[t]
  \centering
    \includegraphics[width=1\linewidth]{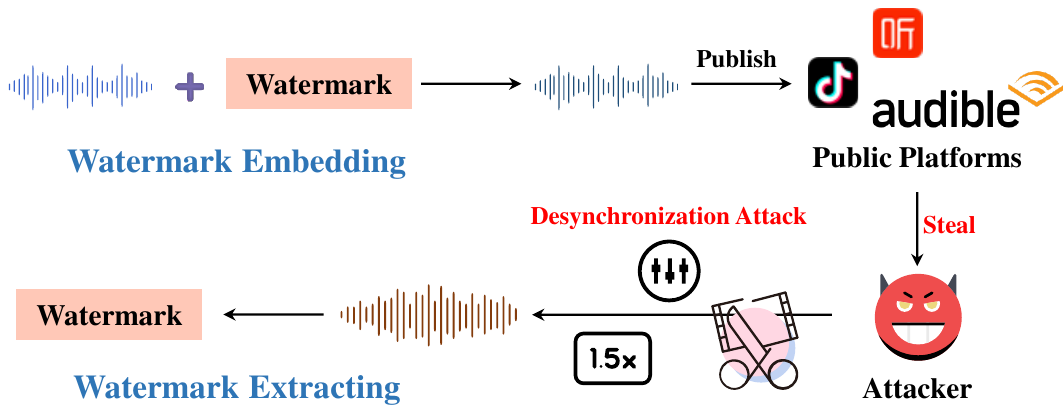}
  \caption{\textbf{The embedding and extraction processes of SyncGuard are meticulously crafted to withstand desynchronization attacks.} }
  \label{attack}
\end{figure}
Recently, pioneering works on audio watermarking based on deep learning have emerged.
Liu et al.~\cite{liu2023dear} propose DeAR, a deep learning-based audio watermarking method. However, it does not address synchronization issues and requires fixed-length input.
Li et al.~\cite{li2024draw} introduce a dual-decoder-based audio watermarking scheme that embeds synchronization codes along with the watermark into audio frames. 
However, when synchronization codes are attacked, it may result in watermark extraction failure.

In this paper, we propose a learning-based scheme named SyncGuard, which uses a frame-wise broadcast embedding strategy to embed the complete watermark information into each audio frame feature, as shown in Fig.~\ref{embedding}(c).
The proposed model enables watermark embedding and extraction in arbitrary-length audio while exhibiting strong resistance to desynchronization attacks.
Specifically, we adopt the classic deep learning-based watermarking framework, which consists of an encoder for embedding the watermark, a distortion layer for simulating potential attacks on the watermarked audio, and a decoder for extracting the watermark.
During watermark embedding, we apply the Short-Time Fourier Transform (STFT) to convert arbitrary-length audio into the frequency domain, obtaining linear spectrograms as carriers for watermark embedding. 
To enhance robustness and reduce dependency on the time domain, the watermark is broadcast at the frame level and integrated with the spectrogram features.
During watermark extraction, the retrieved watermark is averaged along the time dimension. 
The frame-wise embedding strategy eliminates the need to address the localization problem.
To enhance the robustness of the model, our distortion layer incorporates both common signal processing attacks and desynchronization attacks.
Additionally, our encoder and decoder are composed of Dilated Residual (DR) blocks~\cite{yu2017dilated} and Dilated Gated (DG) blocks~\cite{li2018csrnet} to capture multi-resolution time-frequency features more effectively.

Our contributions in this paper can be summarized as follows:
\begin{itemize}
    \item Considering the characteristics of audio signals, we design a frame-wise broadcast embedding strategy to embed the watermark in arbitrary-length audio, eliminating the need for localization during watermark extraction.  
    \item {In SyncGuard}, we employ DR blocks and DG blocks to improve the embedding and extraction capabilities. Additionally, desynchronization attacks are introduced into the distortion layer to enhance the network's robustness.  
    \item {Extensive experimental results} show that our method outperforms existing state-of-the-art audio watermarking approaches in terms of robustness and auditory quality.
\end{itemize}

\section{Related Works}
\subsection{Traditional Audio Watermarking}
Traditional audio watermarking primarily embeds watermark information in the time domain and frequency domain.
Time-domain methods embed watermark by directly altering the values of the audio signal. For example, Xiang et al.~\cite{xiang2011dual} propose a novel dual-channel time-spread echo method for audio watermarking. Compared to time-domain methods, frequency-domain methods can further enhance robustness and achieve better perceptual quality. These methods embed watermark information by slightly adjusting the frequency coefficients using Fourier transforms, such as Discrete Cosine Transform (DCT)~\cite{lei2011blind, liu2021audio} and Discrete Wavelet Transform ~\cite{karajeh2019robust}.

Watermark localization and resistance to desynchronization attacks remain challenging issues. 
To address these challenges, 
Liu et al.~\cite{liu2018patchwork} introduces frequency-domain coefficients logarithmic mean (FDLM) features.
It embeds synchronization codes at the beginning of each audio frame to improve alignment robustness.
However, the decoding process relies on fixed-length segmentation, making it less effective when audio length is altered by attacks such as time-scaling or cropping.
For stereo audio signals, Zong et al.~\cite{zong2020channel} modify the Pearson Correlation Coefficient (PCC) between the DCT coefficients of left and right channel signals. 
Zhao et al.~\cite{zhao2021desynchronization} introduced a novel feature termed Frequency Singular Value Coefficient (FSVC) extracted from DCT domain.
Zhao et al.~\cite{zhao2022ssvs} utilize Segmental Singular Value Summation (SSVS) and Segmental Singular Value Difference (SSVD) features as information carriers and design an adaptive method to select embedding parameters. 
Wu et al.~\cite{wu2025imperceptible} propose LIPAS, leveraging local invariant points and adaptive strength for robust, imperceptible watermarking.
Although these methods can resist desynchronization attacks, the manual design approach limits their flexibility, posing challenges in balancing robustness, audio quality, and embedding capacity.

\subsection{Deep Learning-Based Audio Watermarking}
With the rapid development of deep learning technology, the manual design processes for embedding and extracting audio watermarks are gradually being replaced by neural networks. For example, the Robust-DNN scheme~\cite{pavlovic2022robust} adopts an end-to-end training strategy to learn coefficient embedding methods after STFT. However, this method only focuses on simple attacks, such as Gaussian noise and low-pass filtering, ignoring desynchronization attacks, and is therefore only applicable to fixed-length audio segments.
Inspired by DNN-based image watermarking techniques, Liu et al. propose a deep learning-based watermarking method, DeAR~\cite{liu2023dear}, which injects watermark information into the audio signal in the form of residuals through convolutional neural networks. However, it exhibits limited performance against desynchronization attacks and requires fixed-length audio input.
Recently, Li et al.~\cite{li2024draw} proposed a dual-decoder-based audio watermarking scheme that leverages synchronization codes and learning-enabled techniques to resist desynchronization attacks. 
The method embeds synchronization codes along with the watermark into audio frames and features two independent decoders—one for fixed-length synchronization decoding and the other for variable-length payload decoding. 
However, it may fail when synchronization codes are attacked and shows limited robustness against desynchronization attacks like cropping.  
To address these challenges, we propose SyncGuard, which uses a frame-wise broadcast embedding strategy, eliminating the need for synchronization codes in localization.  

\section{Method}
\begin{figure*}[h]
  \centering
    \includegraphics[width=1.0\textwidth]{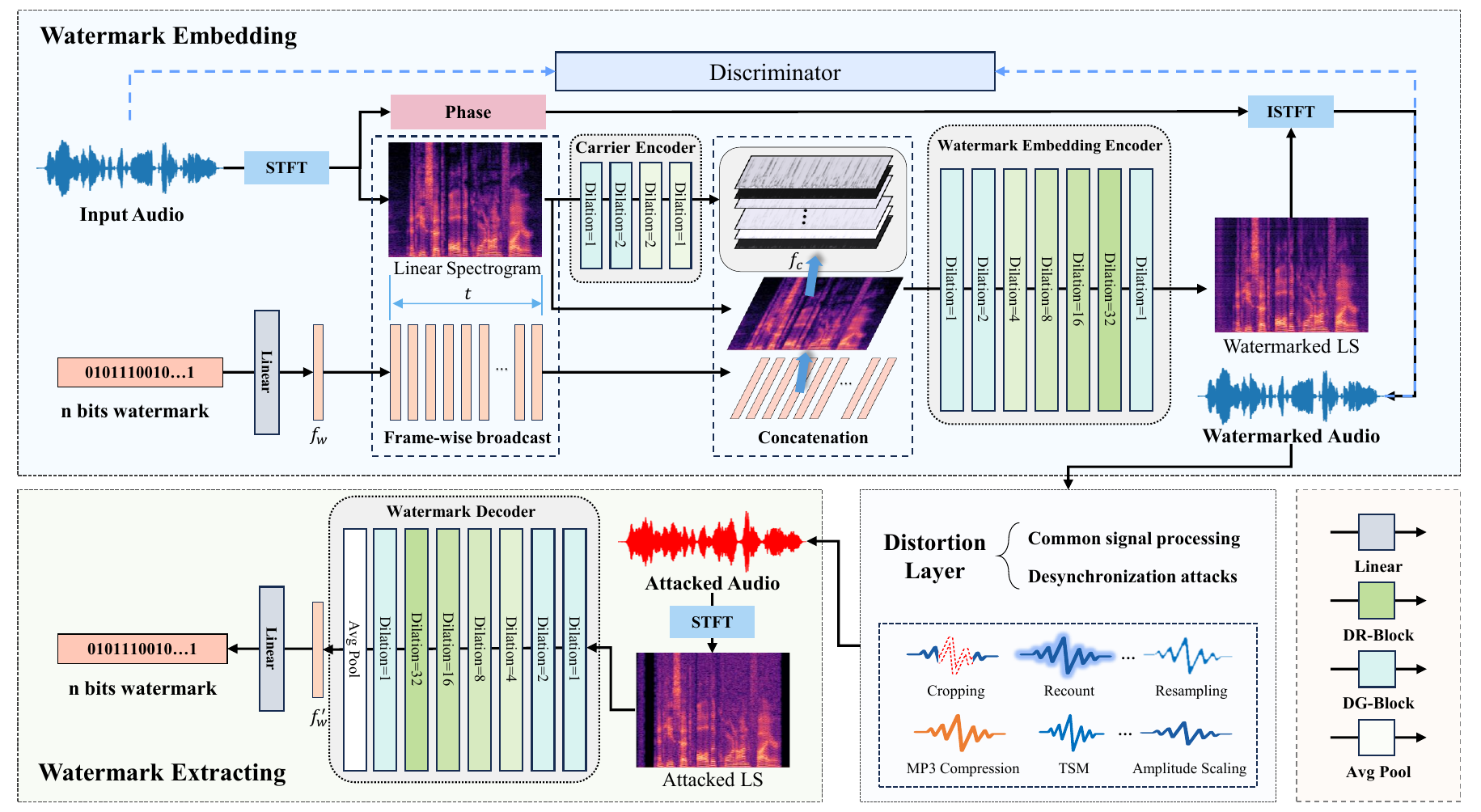}
    \vspace{-7pt}
  \caption{The overview of our training framework. In the Watermark Embedding phase, audio is combined with the watermark message to create watermarked audio, which then undergoes random attacks in a distortion layer. In the Watermark Extracting phase, the watermark is recovered from the distorted audio. }
  \label{main}
\end{figure*}
Fig.~\ref{main} shows the overview of our framework. It consists of three components: a watermark embedding module, a watermark extraction module, and an intervening distortion layer to bolster the robustness against various distortions. Below we provide a detailed description of each component.
\subsection{Watermark Embedding}
Similar to previous audio-based information hiding methods~\cite{pavlovic2022robust,liu2023detecting}, we employ the widely used linear spectrogram of audio as the carrier for watermark embedding. 
Specifically, given single-channel audio $a$ of length $N$, we first apply an STFT operation on it to produce a spectrogram $s$ and the corresponding phase information $p$ as follows:
\begin{equation}
s, p={S T F T}(a) \text{.}
\label{stft}
\end{equation}

We use a linear layer $FC$ to expand the \(n\) bits of the watermark into a vector $f_w$, the same size as one frame of the spectrogram $s$:
\begin{equation}
f_w=FC(w) \text {.}
\end{equation}
\textbf{Frame-wise broadcast.}
Previous methods\cite{liu2023dear,li2024draw} directly extend the watermark vector through linear layers to match the shape of the waveform or features, sacrificing temporal flexibility and leading to a non-uniform distribution of watermark information along the temporal axis.
In our approach, we perform a frame-wise broadcast mechanism of the watermark feature $f_w$ along the time domain until its temporal dimension matches that of $s$, resulting in $f_{wb}$:
\begin{equation}
f_{wb} = \operatorname{Broadcast}\left(f_w, t\right) \text{.}
\end{equation}
This strategy inherently enhances the robustness of watermark information against temporal distortions.

Since the deviation in phase information can severely damage the audio quality, we use the magnitude spectrogram $s$ as the carrier while the phase spectrogram $p$ is only used for signal recovery \cite{ozer2005svd,kreuk2019hide}.
Initially, the magnitude spectrogram \(s\) is fed into the Carrier Encoder $E_{c}$, producing the encoded carrier features \(f_c\):
\begin{equation}
f_c=E_{c}(s)  \text {.}
\end{equation}

Inspired by WaveNet~\cite{oord2016wavenet}, we design the watermark embedding encoder $WE$ using DR and DG blocks to embed the watermark. This design is motivated by the fact that the harmonic intervals in spectrograms are non-uniform and vary with pitch~\cite{wei2022harmof0}. To handle such variability, we employ exponentially increasing dilation rates to construct a hierarchical receptive field, enabling the encoder to capture both local fine-grained details and long-range dependencies across time and frequency.
Next, we concatenate $f_c$, $s$, and $f_{wb}$ to form the input $f_{+}$ for the $WE$:
\begin{equation}
\begin{gathered}
f_{+}=\operatorname{Concatenate}\left(f_c, s, f_{wb}\right) \text {,}\\
s_w={WE}\left(f_{+}\right) \text {.}
\end{gathered}
\end{equation}
Here, $s_w$ represents the watermark embedded spectrogram $s_w$, and
$f_w \in \mathbb{R}^{C_w \times 1 \times H}, f_c \in \mathbb{R}^{C_v \times T \times H}$, and $f_{+} \in$ $\mathbb{R}^{\left(C_w+1+C_v\right) \times T \times H}$.
We employ shortcut connections between different DG blocks, while the watermark and spectrogram are introduced into the network through skip concatenation.
Since the watermark is embedded at the frame level, the frame-wise broadcast module allows information to be embedded across the entire speech signal of any duration, thereby achieving extensive temporal flexibility.

Finally, we reconstruct the watermarked audio \(a_w\) by applying Inverse Short-Time Fourier Transform (ISTFT) to the decoded spectrogram \(s_w\) and the original phase information \(p\):
\begin{equation}
a_w={ISTFT}\left(s_w, p\right)\text{.}
\label{istft}
\end{equation}

\subsection{Watermark Extraction}
Given the watermarked speech $a_w$, the watermark extraction module needs to recover watermark $w^{\prime}$ as consistent as the original watermark $w$. 
In practical applications, during the watermark extraction process, the model slides along the audio, continuously attempting to extract the watermark by combining pattern bits and payloads. The pattern bits are employed as the criterion to validate the correctness of the decoded outputs.
In our end-to-end training process, we focus solely on developing a robust extractor, without performing sliding operations during the extraction phase. Due to our frame-wise broadcasting embedding method, the model inherently exhibits robustness in extracting watermarks from uncertain positions, as will be demonstrated in the experimental section~\ref{flex}.

Initially, we utilize Eq.(\ref{stft}) to perform the STFT on $a_w$ to derive the phase information $p_w$ and spectrogram $s_w$:
\begin{equation}
s_w, p_w={S T F T}(a_w)\text{.}
\end{equation}
$s_w$ is fed into the watermark decoder $DW$ to obtain the recovered watermark feature $f_w^{\prime}$.
The main structure of the decoder is identical to that of $WE$, with an additional average pooling layer along the temporal dimension at the end. Then, by passing it through a linear layer, we can recover the message information:
\begin{equation}
w^{\prime}=FC(DW(s_w)) \text {.}
\end{equation}

To ensure the accuracy of watermark extraction, we introduce a watermark extraction loss $\mathcal{L}_w$, i.e.,
\begin{equation}
\mathcal{L}_w=\frac{1}{N} \sum_{i=1}^N\left(w_i^{\prime}-w_i\right)^2 \text {,}
\end{equation}
where $N$ is the length of the watermark sequence.
\subsection{Imperceptibility Guaranty}
As one of the most important criteria for evaluating digital watermarking, imperceptibility ensures that the embedded watermark cannot be distinguished by human auditory perception.
To ensure the imperceptibility of watermarking, we introduce the watermark embedding loss $L_e$, that is, we adopt the widely-used mean square error MSE as $\mathcal{L}_e$, i.e.,
\begin{equation}
\mathcal{L}_e=\operatorname{MSE}\left(a_w, a\right)=\frac{1}{M} \sum_{i=1}^M\left(\left(a_w\right)_i-a_i\right)^2
\text {,}
\end{equation}
where $M$ is the length of the audio in the time dimension.

To further improve the imperceptibility and minimize the domain gap between $a$ and $a_w$, we adopt the adversarial training strategy to ensure the realism of the generated data, where an extra discriminator $D$ and the adversarial loss $L_{adv}$ is added to make $a_w$ indistinguishable from the pristine $a$, i.e.,
\begin{equation}
\mathcal{L}_{adv}=\log \left(1-\sigma\left({D}\left(a_w\right)\right)\right) .
\end{equation}
Meanwhile, during the training process of ${D}$, $\mathcal{L}_d=\log (1-\sigma({D}(a)))+$ $\log \left(\sigma\left({D}\left(a_w\right)\right)\right)$ is introduced for optimization, where $\sigma(\cdot)$ denotes the sigmoid function.

\subsection{Distortion Layer}
To enhance the robustness of our method against various distortions, we introduce a meticulously designed distortion layer that incorporates both signal processing attacks, such as MP3 compression and Gaussian noise, and desynchronization attacks, including cropping, TSM, and pitch scaling (PS).
This distortion layer processes the watermarked audio signal $a_w$, generating a distorted audio signal $\hat{a_w}$ to challenge the watermark extraction process. 

Ensuring the differentiability of the distortion layer is critical, allowing the model to optimize parameters more effectively in the end-to-end learning process to counteract distortions.
However, desynchronization processes such as TSM and PS are complex and inherently non-differentiable.
To overcome this challenge, inspired by~\cite{kang2010geometric}, we simulate desynchronization attacks as a differentiable process of stretching or compressing the audio signal.
\subsubsection{Time Scale Modification}
The time scale modification attack alters the audio's duration in the time domain while maintaining the pitch. 
Given a watermarked audio $a_w$, the resulting audio after the time scale modification attack, denoted as ${a_{time}}$, is obtained through the distortion layer ${TSM}$, expressed as ${a_{time}}={TSM}\left(a_w\right)$.

To achieve better auditory quality, real-world desynchronization attacks like time scale modification and pitch scale often operate on the audio in the frequency domain. 
Therefore, based on this observation, we utilize the differentiable STFT described in Eq.(\ref{stft}) to obtain the linear spectrogram $s$ and phase $p$ of the audio.
For time scale modification, we linearly stretch or compress the spectrogram $s$ along the time dimension, resulting in $s^{\prime}$:
\begin{equation}
s^{\prime} = Timewarp(s,rate) \text{,}
\label{time}
\end{equation}
where \( \text{rate} \) denotes the time-stretch ratio. The function $Timewarp(\cdot)$ denotes a differentiable interpolation function.
Meanwhile, for each frequency component \( f \), we predict the phase information \( p'_f(t) \) at time \( t \):
\begin{equation}
p'_f(t) = p_f(t) + \Delta p_f(t) \text{.}
\end{equation}
Here, $\Delta p_f(t)$ represents the phase adjustment amount, determined by the phase difference between adjacent time frames and the time scale modification factor $r$ :
\begin{equation}
\Delta p_f(t) = {unwrap}(p_f(t + 1) - p_f(t)) \cdot r\text{.}
\end{equation}
The function ${unwrap}(\cdot)$ is used to handle the periodicity of the phase, ensuring that the phase difference remains within the range of $-\pi$ to $\pi$. Here, $r$ represents the time scale modification factor.
Subsequently, the warped audio signal $a_{time}$ is obtained by applying the $ISTFT$ described in Eq. (\ref{istft}) to the distorted phase and spectrogram:
\begin{equation}
a_{time} = ISTFT(s^{\prime},p^{\prime})\text{.}
\end{equation}

\subsubsection{Pitch Scaling}
The pitch scale attack modifies the pitch of the signal while maintaining its duration. 
Pitch Scale can be simulated by a combination of time scale modification and resampling operations.
Given a watermarked audio \(a_w\), the distortion layer $PS$ induced by pitch scale can be represented as $PS = RS(TSM(a_w))$, where $TSM$ denotes the time scale modification and $RS$ represents resampling.

Specifically, the process begins with the application of time scale modification, where the scaling factor is calculated based on the number of semitones the pitch is shifted:
\begin{equation}
\begin{gathered}
{rate} = 2^{\frac{x}{12}},\\
a_s = Timewarp(a_w,rate)\text{.}
\end{gathered}
\end{equation}
After adjusting the speed, we compute the new sampling rate \(sr\) and employ convolution to resample the audio, restoring it to its original length, obtaining the warped audio signal $a_{pitch}$:
\begin{equation}
a_{{pitch}} = {Conv}(a_s, sr)\text{.}
\label{resample}
\end{equation}

Recognizing that some attacks are inherently more challenging, we have assigned different probabilities to each attack in our scheme.
These probabilities are set at three levels: high (0.3),  medium (0.1), and low (0.05), to reflect the varying degrees of difficulty in achieving robustness across different attack scenarios.
Importantly, this distortion layer is only applied during the training phase and is removed during the actual embedding and extraction operations.

\subsection{Training Strategy}
\label{sec_train}
Simultaneously ensuring accuracy, imperceptibility and robustness is sticky.
Therefore, the training of SyncGuard is divided into two stages.
The first stage only considers the imperceptibility and the accuracy of watermark extraction, 
aiming to build a model that can embed the watermark imperceptibly and extract the watermark accurately.
In the second stage, the requirement for the robustness of watermarking is introduced. The distortion layer is incorporated into the model, and the entire model is trained collectively.
The same total loss function, as introduced in Eq.(\ref{loss}), is applied for both training stages, where $\lambda_e, \lambda_{adv}$, and $\lambda_w$ are hyper parameters of each component.
\begin{equation}
\label{loss}
\mathcal{L}=\lambda_e \cdot \mathcal{L}_e+\lambda_{a d v} \cdot \mathcal{L}_{a d v}+\lambda_w \cdot \mathcal{L}_w\text{.}
\end{equation}

\section{Experiments}
\subsection{Experiment Settings}
\subsubsection{Dataset}
We employ a standard training dataset from LibriSpeech~\cite{panayotov2015librispeech}, which includes audio samples of varying durations, typically around 10 seconds each. For our evaluations, we employ the standard test set from the same dataset, consisting of 2620 audio samples. All audio samples are uniformly resampled to a sampling rate of 22.05 kHz.

\subsubsection{Metrics}
To evaluate the imperceptibility of watermarking, we employ both Signal-to-Noise Ratio (SNR) and Perceptual Evaluation of Speech Quality (PESQ). 
Specifically, while SNR provides a quantitative measure of the quality degradation caused by watermark embedding and audio processing, PESQ offers a more comprehensive assessment by taking into account the characteristics of the human auditory system, making it a superior indicator of speech quality.
Additionally, we assess the robustness of the watermarking schemes using the average bit recovery accuracy (ACC).

\subsubsection{Implementation Details}
In the training process of SyncGuard, we set $\lambda_e=1, \lambda_w=0.01$ and $\lambda_d=0.01$, and utilize Adam~\cite{kingma2014adam} with a learning rate of $10^{-5}$ for optimization by default. 
For STFT, we adopt a filter length of 1024, a hop length of 256, and a window function applied to each frame with a length of 1024.
In the testing process, the embedding capacity is set to 32 bits per second (bps).

\subsubsection{Comparative Methods}
We compared the SyncGuard with state-of-the-art methods, including FSVC~\cite{zhao2021desynchronization}, FDLM~\cite{liu2018patchwork}, DeAR~\cite{liu2023dear} and DRAW~\cite{li2024draw}.
It is important to note that the first two methods do not require any training. 
For DeAR, due to its lack of convergence at high embedding rates, we adhered to its settings and embedded 100 bits.
For DRAW, we followed its settings, embedding at an effective capacity of 32.73 bps, while the embedding capacity for other methods remained at 32 bps. 

\subsection{Experimental Results}
\begin{figure}[!t]
  \centering
    \includegraphics[width=1.0\linewidth]{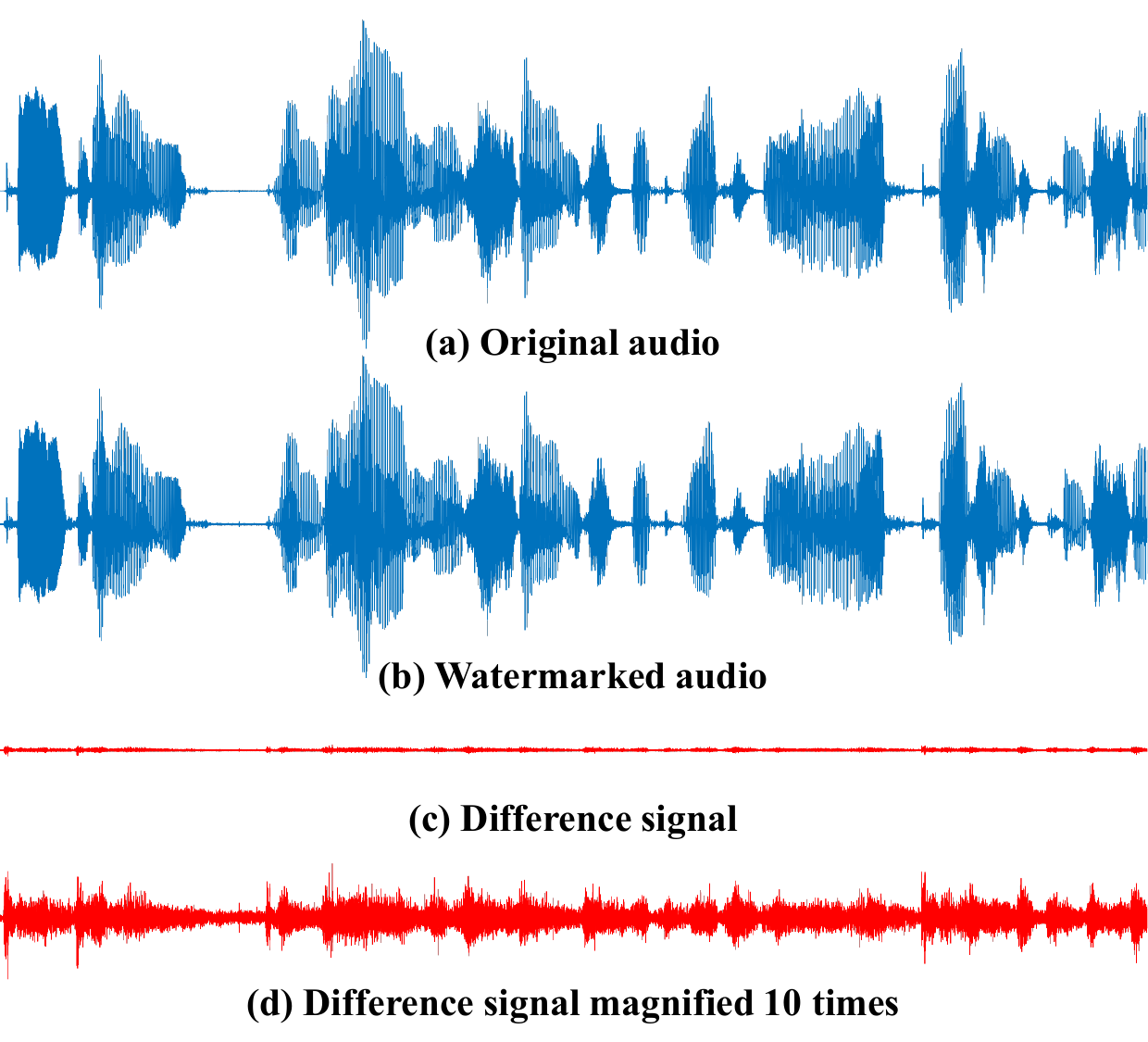}
    \vspace{-15pt}
  \caption{{Visualization of a watermarked audio clip.} }
  \label{wav0}
\end{figure}

\begin{figure}[!t]
  \centering
    \includegraphics[width=1.0\linewidth]{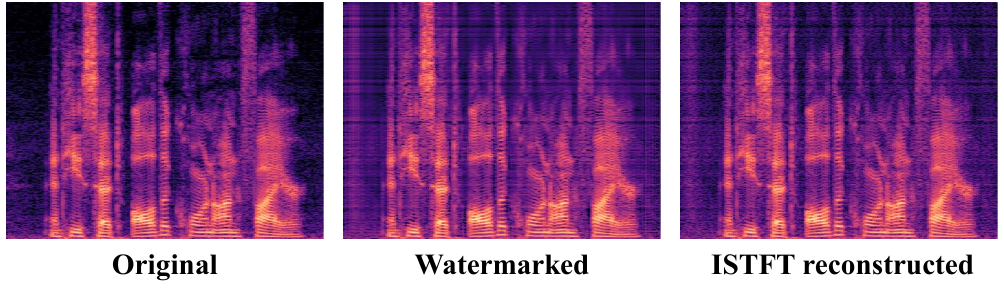}
    \vspace{-15pt}
  \caption{{Visualization of a watermarked audio spectrogram.} }
  \label{ls}
\end{figure}
\subsubsection{Imperceptibility}

\begin{table}[!t]
\centering
\caption{Imperceptibility comparison with the baseline methods.}
\small
{ 
\begin{tabular}{c|ccccc}
\toprule
\textbf{Metrics} & \textbf{FSVC} & \textbf{FDLM} & \textbf{DeAR} & \textbf{DRAW} & \textbf{SyncGuard} \\ 
\midrule
\textbf{SNR} & {{24.23}} & {{25.83}} & {{26.13}} & {{27.17}} & \textbf{{28.27}} \\
\textbf{PESQ} & {{3.72}} & {{3.74}} & {{3.82}} & {{3.93}} & \textbf{{4.02}} \\
\bottomrule
\end{tabular}}
\label{quantitative_comparison} \end{table}

We first evaluate the imperceptibility of the proposed SyncGuard against baseline methods. As indicated in Table~\ref{quantitative_comparison}, SyncGuard achieves the best performance in terms of both SNR and PESQ under similar bps, demonstrating its good imperceptibility.
Additionally, we present visual examples of a watermarked audio and a spectrogram before and after watermark embedding in Fig.~\ref{wav0} and Fig.~\ref{ls}, respectively.
It is apparent that SyncGuard adaptively modifies the original audio.

\subsubsection{Robustness Against Common Signal Processing}
\begin{table}[!t]
\caption{Robustness against common signal processing. * indicates the attack (or corresponding parameter) was not included during training.}
\centering
\setlength{\tabcolsep}{1.0mm}
{
\begin{tabular}{c c | c c c c c}
\toprule
\textbf{Process} & \textbf{Param} & \textbf{FSVC} & \textbf{FDLM} & \textbf{DeAR} & \textbf{DRAW} & \textbf{SyncGuard} \\
\midrule
\multirow{2}{*}{Resample} 
    & 0.8${^{*}}$ & 96.21  & 96.32  & 99.74 & 99.85 & \textbf{99.91} \\
    & 0.9${^{*}}$ & 96.71  & 96.85  & 99.73 & 99.92 & \textbf{100.0} \\
\midrule
\multirow{2}{*}{GS Noise} 
    & 20 dB & 82.03  & 61.89  & \textbf{99.92} & 96.37 & {99.64} \\
    & 30 dB & 90.15  & 62.45  & 99.88 & 99.83 & \textbf{100.0} \\
\midrule
\multirow{1}{*}{MP3} 
    & 64 kbps & 92.11  & 90.37  & 99.37 & 99.53 & \textbf{100.0} \\
\midrule
\multirow{1}{*}{Amplitude} 
    & 85${\%^{*}}$ & \textbf{100.0}  & 98.24  & 99.92 & 99.94 & \textbf{100.0} \\
\midrule
\multirow{1}{*}{Recont} 
    & 8 bps & 97.72  & 75.43  & 99.71 & \textbf{99.91} & {99.82} \\
\midrule
\multirow{1}{*}{Low Pass Filter} 
    & 6kHz${^{*}}$ & 84.83  & 77.25  & 98.62 & {98.32} & \textbf{98.72} \\
\bottomrule
\end{tabular}
}
\label{com}
\end{table}
We have examined the accuracy of our proposed model under various common signal processing distortions.
As indicated in Table~\ref{com}, our model consistently demonstrates superior performance across various attacks, exhibiting strong robustness even when confronted with unseen attacks during training.

\begin{table}[!t]
\caption{Robustness against desynchronization attacks. * indicates the attack (or corresponding parameter) was not included during training.}
\centering
\setlength{\tabcolsep}{1.5mm}
{ 
\begin{tabular}{c c|c c c c c}
\toprule
\textbf{Attack} & \textbf{Param} & \textbf{FSVC} & \textbf{FDLM} & \textbf{DeAR} & \textbf{DRAW} & \textbf{SyncGuard} \\
\midrule
\multirow{1}{*}{Jittering${^{*}}$} 
    & 1/100 & 79.81  & 82.63  & 99.82 & {99.93} & \textbf{100.0} \\
\midrule
\multirow{4}{*}{TSM} 
    & 0.8 & 50.91  & 56.38  & 49.47 & {80.83} & \textbf{97.72} \\
    & 0.9 & 51.32  & 51.91  & 55.87 & {95.36} & \textbf{100.0} \\
    & 1.1 & 50.61  & 57.18  & 53.29 & {99.05} & \textbf{100.0} \\
    & 1.2 & 50.72  & 50.34  & 47.72 & {98.13} & \textbf{98.36} \\
\midrule
\multirow{2}{*}{Cropping} 
    & 10\%${^{*}}$ & 49.33  & 50.78  & 79.54 & {89.62} & \textbf{100.0} \\
    & 20\%${^{*}}$ & 51.24  & 50.61  & 62.73 & {79.37} & \textbf{100.0} \\
\midrule
\multirow{2}{*}{PS} 
    & 0.9 & 51.42  & 50.73  & 62.08 & {98.04} & \textbf{99.92} \\
    & 1.1 & 47.25  & 50.41  & 61.89 & {99.60} & \textbf{99.83} \\
\bottomrule
\end{tabular}}
\label{de}
\end{table}

\subsubsection{Robustness Against Desynchronization Attacks}
We employ the following desynchronization attacks to assess the robustness of our proposed method and the methods chosen for comparison:
\textbf{(1) Jittering attack}: The watermarked audio are randomly deleted one sample from every 100 consecutive samples.
\textbf{(2) Cropping}:The watermarked audio was cropped at random positions, with 10\% and 20\% of the audio removed.
\textbf{(3) TSM}: Time scale the watermarked audio with attack factors of $0.8,0.9,1.1$, and $1.2$.
\textbf{(4) PS}: Pitch scale the watermarked audio with attack factors of $0.9$ and $1.1$.

As shown in Table~\ref{de}, our method demonstrates strong robustness against all the aforementioned desynchronization attacks. 
Notably, SyncGuard demonstrates exceptional performance in resisting severe TSM, achieving over 97\% extraction accuracy across four parameter settings. 
Additionally, it significantly outperforms other methods in handling cropping, achieving 100\% extraction accuracy even when 20\% of the audio is cropped, while other methods fall below 80\% accuracy under the same conditions.
\begin{figure}[t]
  \centering
    \includegraphics[width=0.9\linewidth]{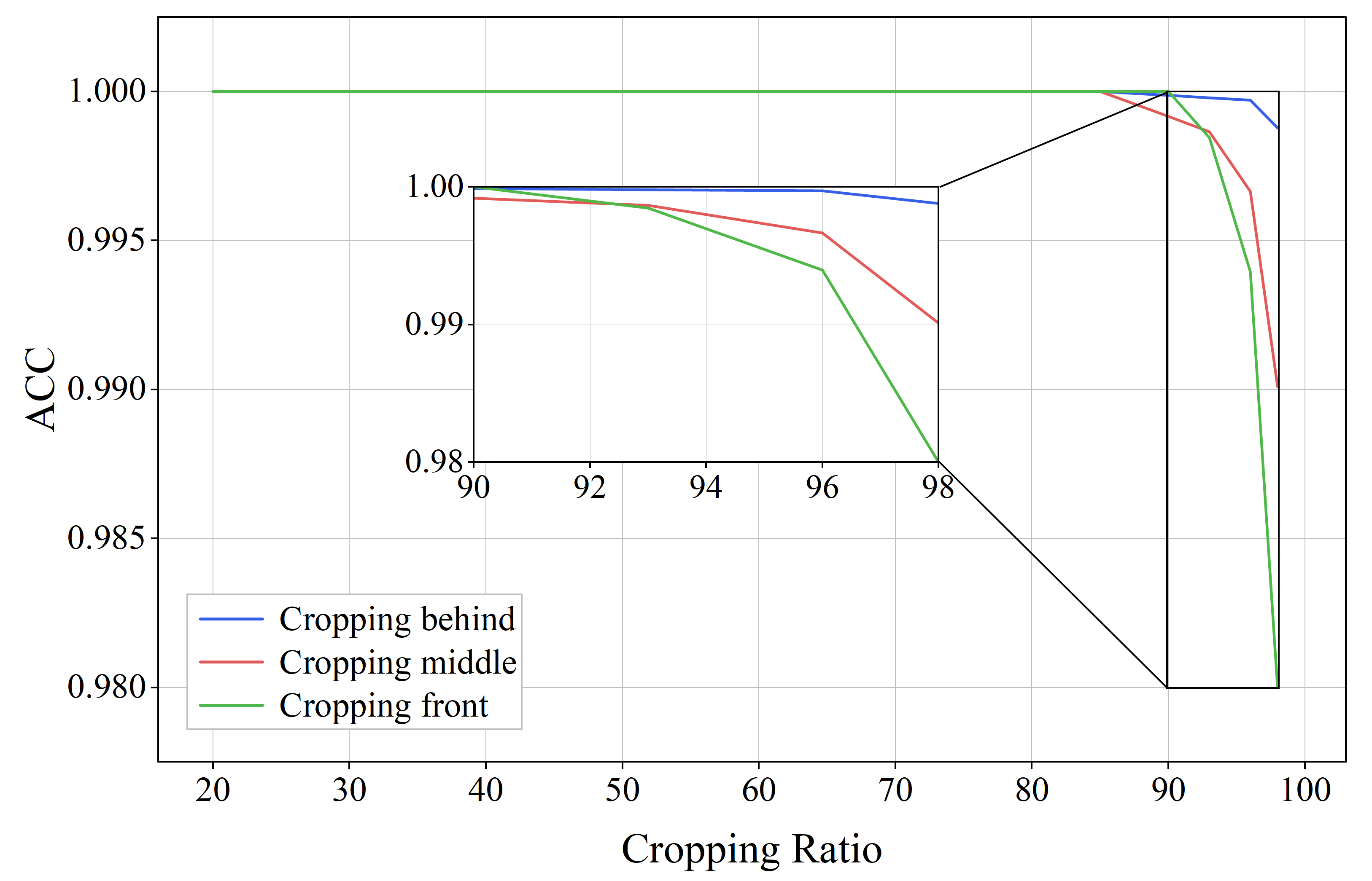}
  \caption{{Robustness against different cropping strategies.} }
  \label{crop}
\end{figure}

To further investigate SyncGuard's robustness against cropping, we conducted tests by cropping the watermarked audio at three different positions: the beginning, middle, and end, and assessed the ACC under varying cropping ratios. As depicted in Figure~\ref{crop}, even with 85\% of the audio cropped, our method maintains a 100\% accuracy.
This superior performance is attributed to our frame-wise broadcast embedding strategy, which ensures that the embedded watermark is uniformly and comprehensively distributed across all parts of the audio, thereby providing strong resilience against cropping.

\subsubsection{Flexibility}
\label{flex}
We evaluated the flexibility of our method in both watermark embedding and extraction processes. Specifically, we embedded watermarks in audio segments with minimum durations of 0.5s, 1s, 5s, and 10s, and measured the ACC after subjecting them to Gaussian noise distortion. As shown in Table~\ref{length}, variations in the minimum unit length did not lead to a decrease in performance.

\begin{figure}[h]
  \centering
    \includegraphics[width=1.0\linewidth]{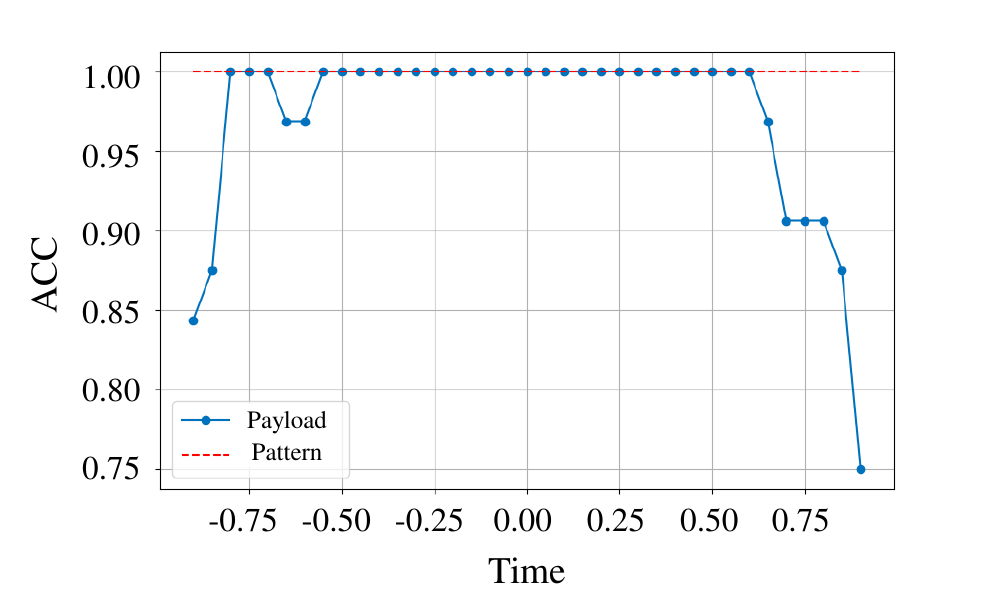}
    \vspace{-15pt}
  \caption{{Performance at different time offsets from the watermark insertion position. Positive values on the time axis indicate a rightward shift, while negative values indicate a leftward shift.} }
  \label{acc_shift}
\end{figure}
During the extraction phase, we investigated the effectiveness of our proposed method in watermark localization. We inserted a 1-second watermark at a random position within a 3-second audio segment and then performed decoding. The model continuously attempts to extract the watermark by sliding a 0.05-second window across the audio.
The results in Fig.~\ref{acc_shift} show that our method can accurately extract the watermark within a range of 0.5 seconds of offset from the original watermark position.
\begin{table}[!t]
\caption{Performance across different minimum audio segment lengths.}
\centering
\small
{ 
\begin{tabular}{c|cccc}
\toprule
\textbf{Metrics} & \textbf{0.5s} & \textbf{1s} & \textbf{5s} & \textbf{10s} \\ 
\midrule
\textbf{ACC} & {{99.63}} & \textbf{{100.0}} & \textbf{{100.0}} & {{99.74}}\\
\textbf{SNR} & {{28.67}} &{{28.37}} &\textbf{{29.34}} &{{28.93}}\\
\textbf{PESQ} & {{3.93}} & {{3.92}} & {{3.98}} & \textbf{{4.01}}\\
\bottomrule
\end{tabular}}
\label{length} \end{table}

\begin{figure}[!t]
  \centering
    \includegraphics[width=0.95\linewidth]{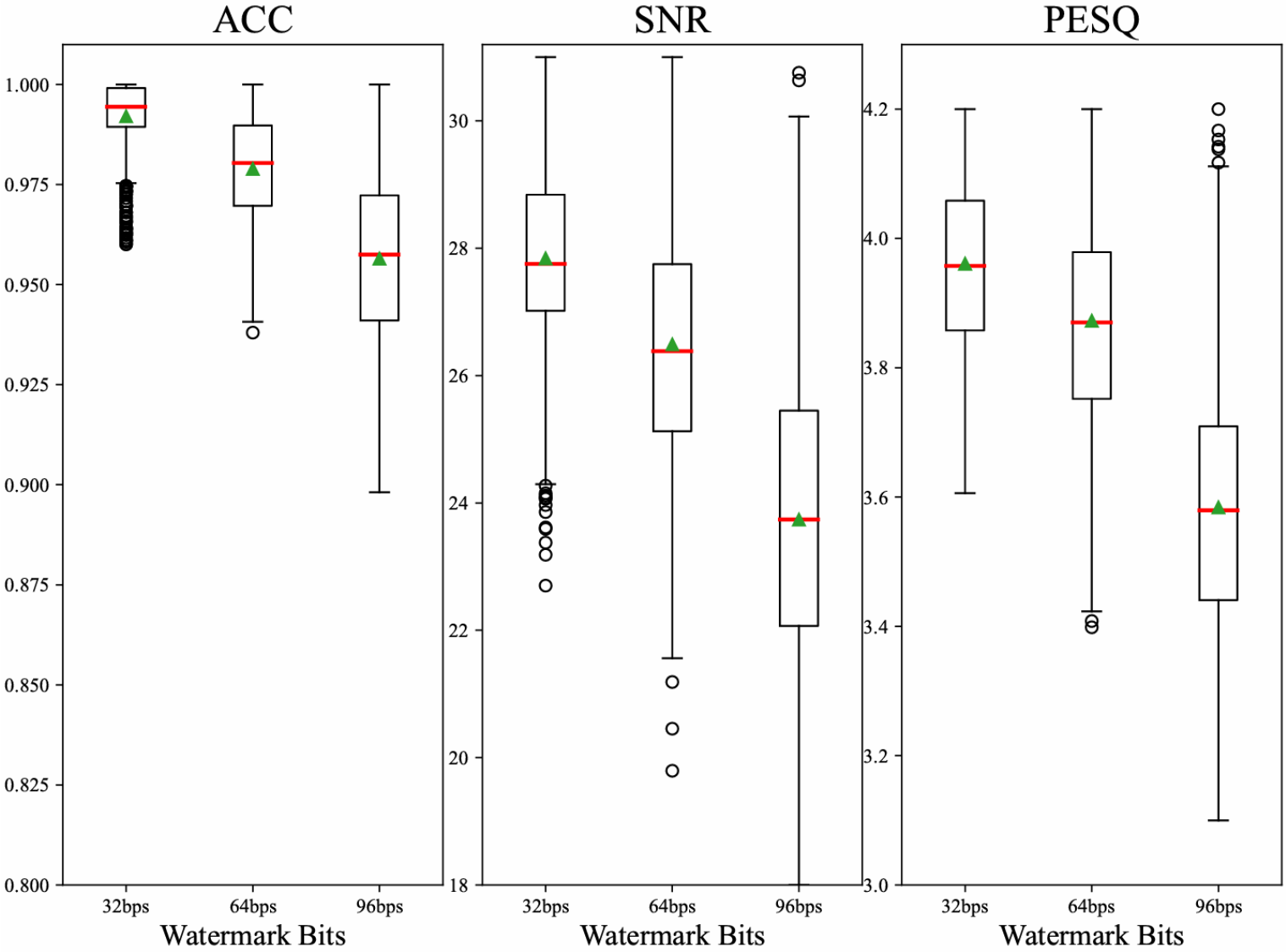}
  \caption{{Performance under different watermark bits.} }
  \label{acc_snr}
\end{figure}

As shown in Table~\ref{eff}, the encoder has 0.70M parameters and 44.42 GFLOPs, with an average inference time of 3.34 ms per second of audio. These results demonstrate that SyncGuard is efficient enough for real-time or edge deployment scenarios.

\begin{table}[!t]
\centering
\caption{Efficiency analysis of SyncGuard.}
\small
{
\begin{tabular}{c|ccc}
\toprule
\textbf{Module} & \textbf{Params (M)} & \textbf{FLOPs (G)} & \textbf{Time (ms / sec)} \\
\midrule
\textbf{Encoder} & 0.70 & 44.42 & 3.34 \\
\textbf{Decoder} & 0.41 & 24.68 & 1.17 \\
\bottomrule
\end{tabular}}
\label{eff}
\end{table}

\subsection{Ablation Study}
\subsubsection{Influence of Watermark Bits}
We conducted a study to evaluate the influence of different watermark bits.
Specifically, we trained three model variants with 32, 64, and 96 bit watermarks and evaluated the quality and Gaussian-noise robustness of the watermarked audio.
As illustrated in Fig.~\ref{acc_snr}, when increasing the watermark bits from 32 to 64 bps, the ACC remains stable around 100\%, and the SNR hovers around 27 dB. However, a noticeable decline in both metrics is observed when the watermark bits is increased from 64 to 96 bps.

\subsubsection{Importance of DR blocks and DG blocks}
\begin{figure}[!t]
  \centering
    \includegraphics[width=1.0\linewidth]{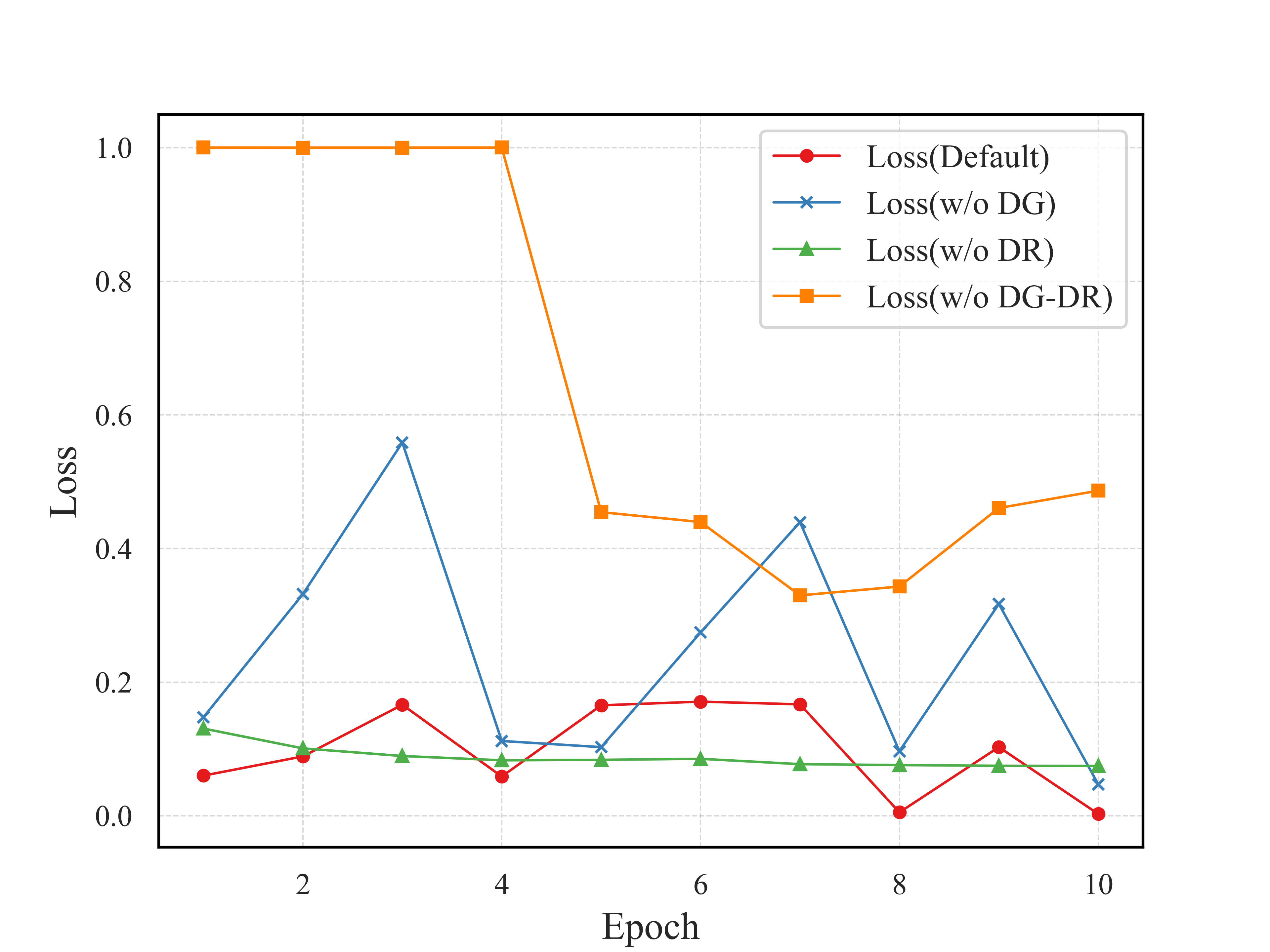}
    \vspace{-7pt}
  \caption{{The Loss vs. Epoch of the First-Stage training.} }
  \label{loss_epoch}
\end{figure}
A key advantage of SyncGuard lies in its use of DR and DG blocks. To demonstrate their effectiveness, we conducted an ablation study comparing the original SyncGuard with a modified version where DR or DG blocks were replaced by 2D convolutional networks, as shown in Fig.~\ref{loss_epoch}. The results indicate that models with DR or DG blocks achieve rapid convergence within the first epoch, whereas models using 2D convolutional networks struggle to converge.
Specifically, DG blocks exhibit stable convergence but are prone to premature convergence to local optima. In contrast, DR blocks tend to exhibit minor oscillations and jumps after convergence. By integrating the strengths of both blocks, we achieve a balance between stability and flexibility, preventing the model from prematurely settling into local optima while enabling robust convergence under complex distortion layers. 
\subsubsection{Importance of Distortion Layer}
%
\begin{table}[!t]
\caption{The robustness performance (ACC) of different
configurations.}
\centering
\small
{
\begin{tabular}{c c | c c c}
\toprule
\textbf{Corruption} & \textbf{Parameter} & \textbf{Default} & \textbf{w/o TSM} & \textbf{w/o PS} \\ 
\midrule
\multirow{1}{*}{Jittering} 
    & 1/100 &  \textbf{100} &  99.13 &  99.18 \\
\midrule
\multirow{2}{*}{TSM} 
    & 0.9 &  \textbf{100} &  90.04 &   100\\
    & 1.1 &  \textbf{100} &  91.38 &   100\\
\midrule
\multirow{2}{*}{PS} 
    & 0.9 &  \textbf{99.92} &  95.32 &   61.32\\
    & 1.1 &  \textbf{99.83} &  94.37 &  59.64\\
\bottomrule 
\end{tabular}}
\label{de}
\end{table}
To thoroughly assess the contribution of TSM and PS distortions, we retrained the model twice, each time excluding one type of distortion. As shown in Table~\ref{de}, the two distortion types exhibit a mutually reinforcing effect. In particular, PS distortion plays a pivotal role, yielding a substantial accuracy gain of 38.60\% under PS attacks. Since PS inherently includes TS-related processes, training with PS distortion alone still provides effective defense against TSM attacks. In contrast, using only TSM distortion is insufficient to counter PS attacks. Notably, although TSM is a subset of PS, incorporating TSM distortion still leads to additional performance gains, improving accuracy by 4.60\% under PS attacks and 9.96\% under TSM attacks.

\section{Conclusion}
In this paper, we introduce SyncGuard, a robust audio watermarking scheme designed to counteract desynchronization attacks. 
SyncGuard employs a frame-wise broadcast embedding strategy, enabling watermark embedding and extraction in arbitrary-length audio while demonstrating strong resistance to desynchronization attacks. 
The frame-wise embedding strategy also eliminates the need to address localization issues.
To further enhance robustness against various distortions, we introduce a meticulously designed distortion layer. 
Additionally, to improve embedding and extraction capabilities, we integrate DR and DG blocks within the framework.
Experimental results show that our method achieves satisfactory performance from a comprehensive perspective of robustness and imperceptibility.

\begin{ack}
This work was supported by the Natural Key R\&D Program of China under Grant 2023YFF0905000.
\end{ack}



\bibliography{m8910}

\end{document}